\documentclass[conference]{IEEEtran}
\IEEEoverridecommandlockouts
\usepackage{cite}
\usepackage{amsmath,amssymb,amsfonts}
\usepackage{algorithmic}
\usepackage{graphicx}
\usepackage{textcomp}
\usepackage{xcolor}
\def\BibTeX{{\rm B\kern-.05em{\sc i\kern-.025em b}\kern-.08em
    T\kern-.1667em\lower.7ex\hbox{E}\kern-.125emX}}

\usepackage[pagebackref=true,breaklinks=true,colorlinks,bookmarks=false]{hyperref}

\begin{document}

\title{WiCluster: Passive Indoor 2D/3D Positioning using WiFi without Precise Labels}

\author{Ilia~Karmanov, Farhad~G.~Zanjani, Simone~Merlin, Ishaque~Kadampot, Daniel~Dijkman\\
{Qualcomm AI Research\thanks{ Qualcomm AI Research is an initiative of Qualcomm Technologies, Inc}}\\
{\tt\small \{ikarmano,fzanjani,ikadampo,smerlin,ddijkman\}@qti.qualcomm.com}}



\maketitle

\begin{abstract}

We introduce WiCluster, a new machine learning (ML) approach for passive indoor  positioning  using  radio  frequency  (RF)  channel  state  information (CSI). WiCluster can predict both a zone-level position and a precise 2D or 3D position, without using any precise position labels during training. 
Prior CSI-based indoor positioning work has relied on non-parametric approaches using digital signal-processing (DSP) and, more recently, parametric approaches (e.g., fully supervised ML methods). However these do not handle the complexity of real-world environments well and do not meet requirements for large-scale commercial deployments: the accuracy of DSP-based method deteriorates significantly in non-line-of-sight conditions, while supervised ML methods need large amounts of hard-to-acquire centimeter accuracy position labels.
In contrast, WiCluster is precise, requires weaker label-information that can be easily collected, and works well in non-line-of-sight conditions. 
Our first contribution is a novel dimensionality reduction method for charting. It combines a triplet-loss with a 
multi-scale clustering-loss to map the high-dimensional CSI representation to a 2D/3D latent space. Our second contribution is two weakly supervised losses that map this latent space into a Cartesian map, resulting in meter-accuracy position results. These losses only require simple to acquire priors: a sketch of the floorplan, approximate access-point locations and a few CSI packets that are labeled with the corresponding zone in the floorplan.
Thirdly, we report results and a robustness study for 2D positioning in two single-floor office buildings and 3D positioning in a two-story home.\end{abstract}


\section{Introduction}
\begin{figure}[htbp]
\centerline{\includegraphics[width=1.0\linewidth]{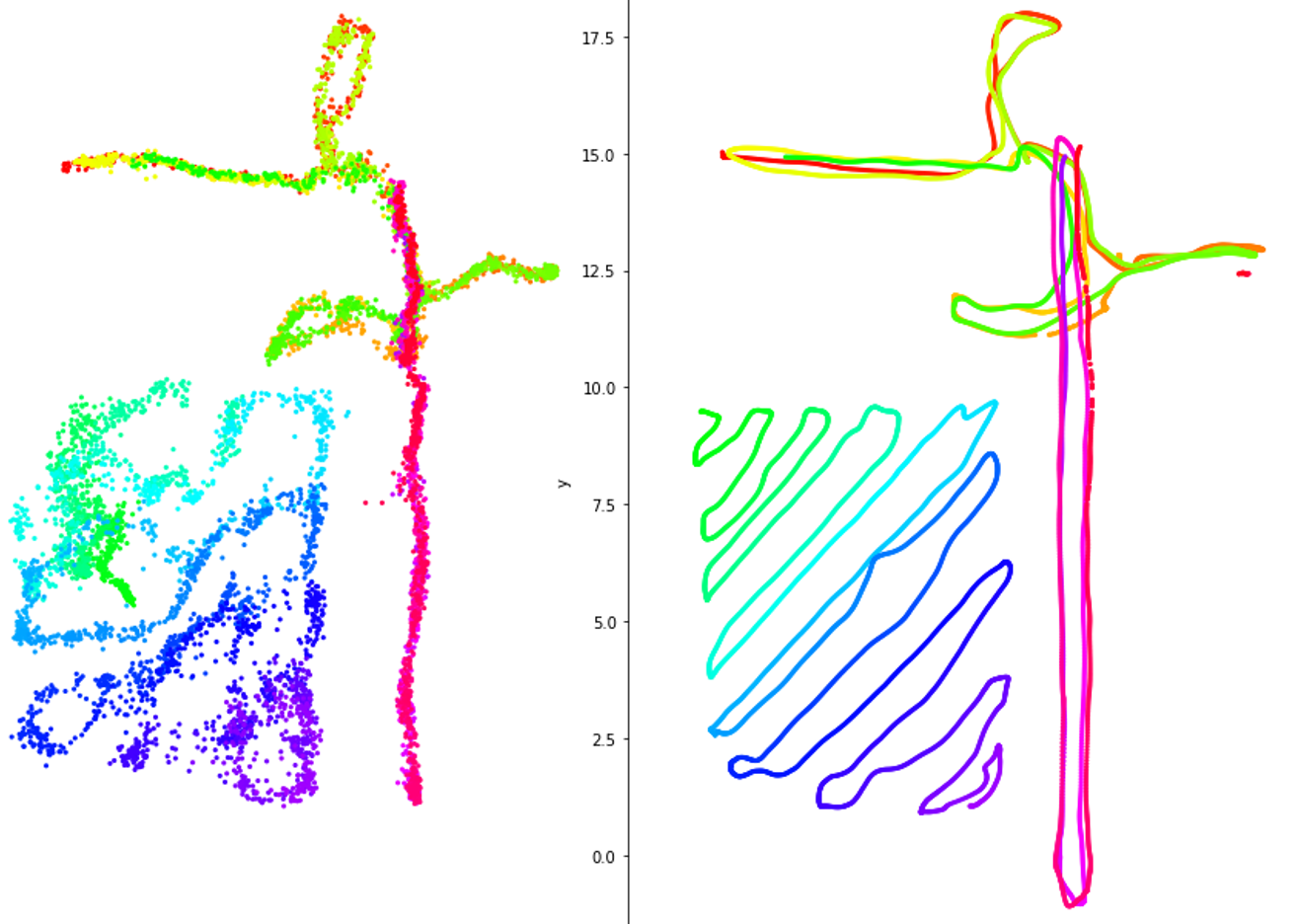}}
\caption{\textbf{WiCluster:} Prediction (left) vs Ground-Truth Locations (right). Color-scale corresponds to the timestamp of the CSI samples. Scale is in meters.   
}
\label{fig:prediction_train}
\end{figure}

We introduce WiCluster, a new algorithm that uses self-supervision to learn a model that predicts the precise position of a person in an indoor-environment using channel state information (CSI) from WiFi access points. We specifically address a real-world deployment use-case, where precise position labels will not be available (since ground truth data-collection is too cumbersome, expensive, and privacy sensitive for consumers) and access points may be in non-line-of sight conditions with respect to each other and with respect to the target (e.g. in different rooms). 
Our solution generates a topologically accurate 2D or 3D latent-space, which we transport into the real-world space by making use of priors such as the floor-plan of the building, anchors such as locations of the access-points, and landmarks which are a small set of room-level annotations. We predict a precise position for a single person in a multi-room environment, without using any precise position labels during training, as shown in Fig.~\ref{fig:prediction_train}. 

Passive positioning, unlike active positioning, does not involve the participation of the target in the positioning problem, i.e. the target is not required to carry a positioning device such as a mobile phone. It exploits the perturbations in the propagation of RF signals between a transmitter and a receiver, neither of which are carried by the target, and operates like a bi-static radar. Passive positioning using WiFi can be useful in the domain of home, enterprise and industrial automation and robotics. Zone-level positioning refers to the ability to determine whether the target is present in a certain area of a building, such as a room. It can enable, for instance, intrusion detection, smart energy usage, building space usage optimization, etc.

Precise positioning refers to the ability to determine the 2D/3D coordinates of the target within the environment; in this paper we target sub-meter level positioning which can enable, for instance, asset and customer tracking in a business environment, indoor navigation, etc. It is not affected by poor light conditions and can thus work in the dark, and it can work across walls. Also, a device without a video-camera is more likely to be adopted by a user since it largely mitigates issues of privacy related to video images.
Passive positioning, and in general RF sensing, is also the subject of the IEEE 802.11bf Task Group, which is currently defining the standard signalling to enable inter-operation across WiFi sensing devices. Such standardization will enable the growth of an ecosystem of sensing devices which will further expand the applicability of the solutions described in this paper. 

We build upon Ferrand et al, \cite{WirelessChannelCharting}, who use a triplet-loss to train a neural-network in an unsupervised manner to simultaneously learn a spatial similarity metric between CSI samples and to perform dimensionality reduction to a 2D latent space that is topologically close to the true geographic environment. We find that by combining a triplet-loss with a cluster prediction loss, we are able to improve upon the 2D latent-space: \textbf{the triplet-loss encourages the model to learn a representation that brings together CSI samples that are close in time (and thus in space), and the clustering extends this to CSI samples that are close in space but not necessarily close  in time.}

\begin{figure}[htbp]
\centerline{\includegraphics[width=1.0\linewidth]{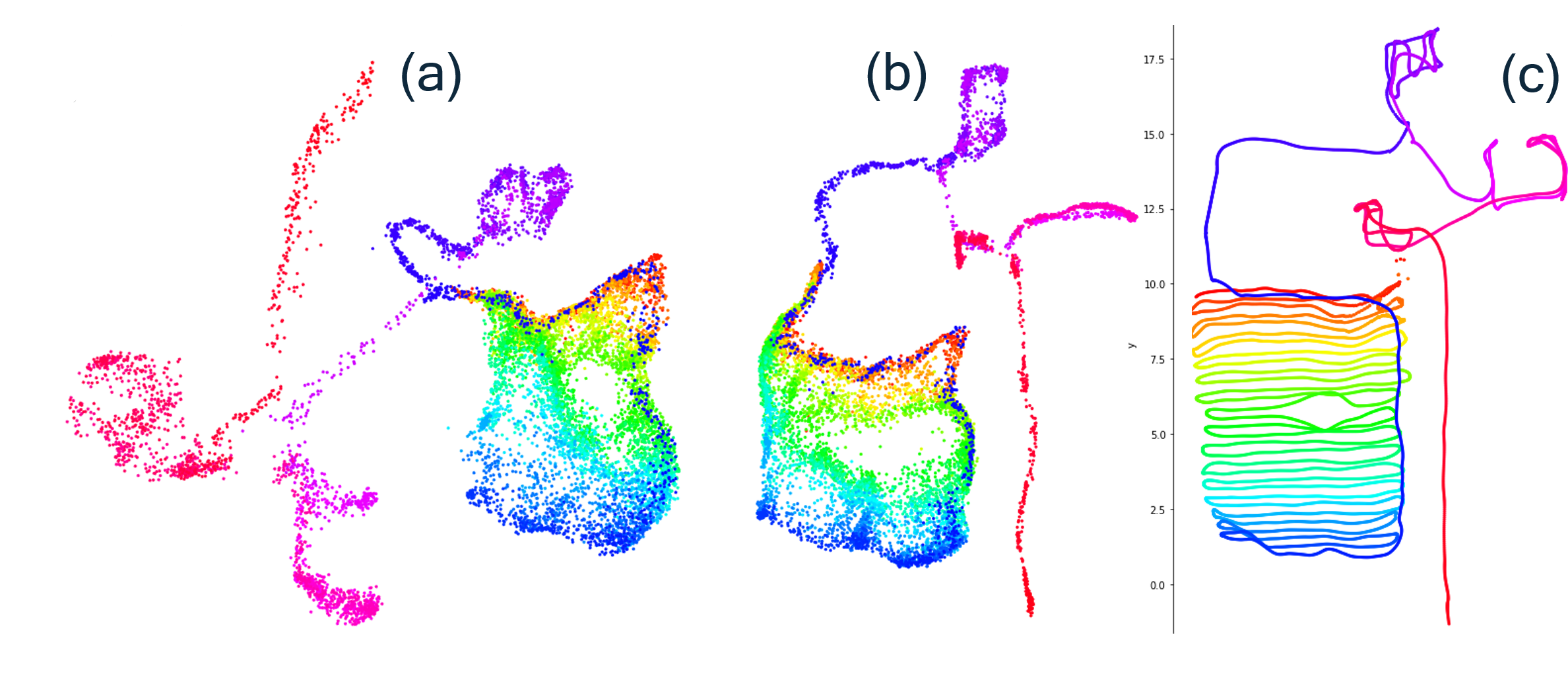}}
\caption{\textbf{(a) Unsupervised 2D latent-space, (b) Weakly-supervised Cartesian map and (c) the ground-truth.} By incorporating a few room-level labels with a floor-plan and introducing the location of the access-points, the unsupervised latent-space is aligned with the floor-plan.}
\label{fig:latent_and_cartesian}
\end{figure}

Our contributions are: (i) a novel dimensionality  reduction  method for charting. It combines representation learning with a new dimensionality reduction technique in order to embed a high-dimensional representation of the CSI in a 2D latent space, as shown in Fig.~\ref{fig:latent_and_cartesian}a. The dimensionality reduction approach preserves cluster membership across dimensions at multiple scales to preserve local and global structure within the data. (ii) We introduce two additional weakly-supervised losses (a zone loss and an access-point loss) to transport this latent space into the real-world space, as shown in Fig.~\ref{fig:latent_and_cartesian}b, by incorporating some priors in our model. (iii) We evaluate our model and demonstrate state-of-the-art results on three different data-sets that are specifically designed to mimic an actual real-world deployment use-case (i.e. not a simple lab environment). 

\section{Related Work}

Our work builds on both wireless indoor positioning and representation learning.

\textbf{Classical DSP} methods for indoor positioning \cite{mdtrack, widar20} can determine the location of subjects by assuming that the perturbation of the RF signal propagation, induced by the target motion, follows a known mathematical model. In reality, such models work when the target is within line-of-sight of the transmitter and receiver, but fail in environments with non-line-of-sight propagation and with complex reflection patterns.

\textbf{Supervised Learning} \cite{piw, skeleton, csinet}  has been shown to accurately predict location from CSI, however it requires a large amount of precise annotations, which cannot be practically obtained by a user deploying the system in, e.g., a  home. Our method instead requires only a few room-level labels (per room), which can be easily collected with a mobile-app.

\textbf{Manifold Learning} techniques project the high-dimensional CSI signal into a 2D latent-space that preserves some notion of similarity. These can be typically divided into distance-preserving and neighbour-preserving methods. Distance-preserving methods include Multidimensional Scaling~\cite{mds},  \cite{kruskal} which tries to preserve all pairwise (Euclidean) distances between the embedding and the original, high-dimensional space. This has been extended to methods like Isomap~\cite{isomap}, which preserve an estimate of a geodesic distance by creating a shortest-path through points on the manifold and more recently L-Isomap~\cite{l_isomap}, which reduces the complexity by selecting landmark points. Neighbour-preserving methods include t-SNE~\cite{tsne} which optimizes the Kullback–Leibler divergence between probability distributions across the two spaces and UMAP~\cite{umap} which optimizes the fuzzy-set binary cross-entropy. Typically the neighbour-preserving approaches will preserve local structure and distance-preserving approaches will preserve global structure~\cite{pymde}.

\textbf{Contrastive Learning} methods can also create a 2D-embedding by enforcing that points should be closer to their respective positive sets than their negative sets. Approaches like Ivis~\cite{ivis} and TriMap \cite{trimap} use methods like k-nearest neighbour (KNN) to determine the sets of positive and negative points sampled for a triplet-loss. Other approaches like Ferrand et al \cite{WirelessChannelCharting} make use of the observation that the data is sequential and points closer in time should also be closer in the 2D latent-space. Although we also use such a loss, it is part of a larger model and our  experiments show that by itself it does not capture global structure in our data well, perhaps because this approach does not operate on the full population of samples but on a batch-level.

\textbf{Neural Clustering} inspired by DeepCluster~\cite{DeepCluster} has been used for self-supervised learning to create a high-dimensional embedding that can be fine-tuned in a fully-supervised manner for downstream tasks (e.g. image classification). However, the novelty in our approach lies in creating a 2D-embedding, using a cross-dimension architecture, not just a high-dimension feature vector. Alwassel et al~\cite{xdc} have proposed an architecture to train on videos by crossing the appearance and audio modality, however whereas they operate cross-modality we operate cross-dimensions. Coron et al~\cite{swav} also use clustering; however instead of crossing the prediction between different dimension-projections they use different augmentation-views of the same image.

\section{Method}

\subsection{Background}

\textbf{DeepCluster}: The cluster component of our model is inspired by DeepCluster~\cite{DeepCluster}. For convenience, we describe it shortly in this section. DeepCluster alternates between (i) k-means clustering the convolutional neural-net (convnet) representation, parameterized by $\theta$, of the input-data: $f_\theta(x_i)$ to create pseudo-labels using 
\begin{equation*}
\label{equation:kmeans}
  \min_{C \in \mathbb{R}^{d\times k}}
  \frac{1}{N}
  \sum_{n=1}^N
  \min_{y_n \in \{0,1\}^{k}}
  \| f_\theta(x_n) -  C y_n \|_2^2
  \quad
  \text{s.t.}
  \quad
  y_n^\top 1_k = 1
\end{equation*}
and then (ii) back-propagates the cross-entropy loss 
\begin{equation}
\mathcal{L_C}
=
-
\frac{1}{N}
\sum_{i=1}^N
\log p(y_i|x_i),
\label{equation:ce}
\end{equation}
for each batch of length $N$, when the (same) convnet is trained to predict those pseudo-labels. A pseudo-label is a one-hot cluster assignment $y_n$ for each data-point, which corresponds to the the centroid matrix $C \in \mathbb{R}^{d\times k}$, where we learn $k$ centroids with dimensionality $d$. 

Taking the cross-entropy loss as given, the ideal set of discrete position pseudo-labels would partition the plane into regions that resemble Voronoi cells. This is exactly what k-means tries to create and why w\textbf{}e use it instead of other methods such as spectral clustering.

\begin{figure}[tbp]
\centerline{\includegraphics[width=1.0\linewidth]{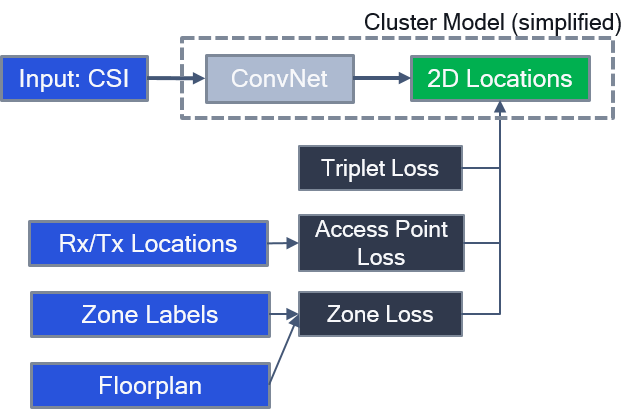}}
\caption{\textbf{Full Architecture} - WiCluster takes CSI as input and uses neural clustering to produce a latent space of 2D locations. Triplet loss enhances the quality of this latent space. Two additional losses embed the latent space into a Cartesian map / floor-plan.}
\label{fig:arch_a}
\end{figure}

\subsection{Learning the 2D latent space (self-supervised)}

\begin{figure}[tbp]
\centerline{\includegraphics[width=1.0\linewidth]{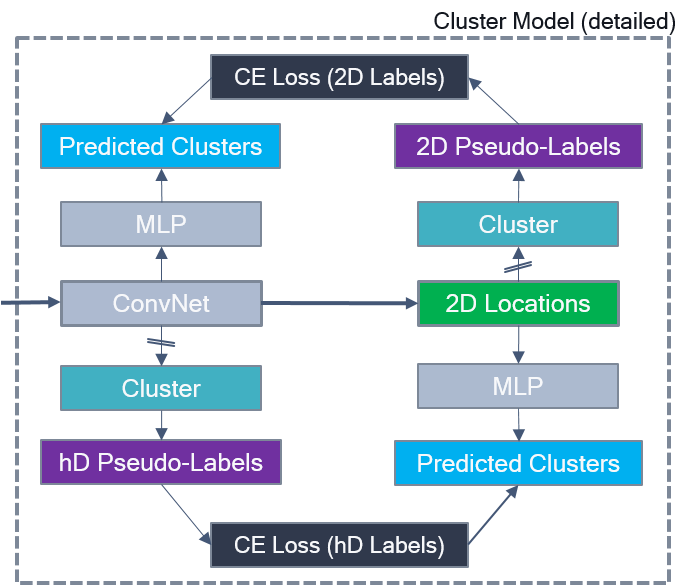}}
\caption{\textbf{Cluster Model} - the high-dimensional features from the convnet are clustered to generate a set of pseudo-labels. The 2D latent-space features are also clustered to generate another set of pseudo-labels. We cross the targets and predictions so that the MLP trained on 2D-projected features predicts clusters obtained from the high-dimensional features, and vice versa. There is no gradient flow for clustering, illustrated by the stop-arrows.}
\label{fig:arch_b}
\end{figure}

\textbf{Cluster-loss:} We introduce a way to perform dimensionality reduction using clustering. First, we extract cluster assignments from a high-D representation of the data. The high-D representation is then projected to a 2D latent-space. A small Multilayer Perceptron (MLP) is attached to the 2D latent-space and the network is trained to predict the cluster assignments with a cross-entropy loss. This results in a 2D latent-space that brings together points that belong to the same cluster in the high-D space and separates them from points that belong to a different cluster. 

We observe that optimizing the high-D representation indirectly through only the 2D representation would often get stuck in a local minima, owing to the mapping between the two spaces not being bijective\cite{knot}. To help the 2D projection unfold the high-D manifold we also allow the network to directly optimize the high-D space: we attach an MLP to the high-D features and try to predict cluster assignments obtained from the 2D projection. Hence, the main component of our model is a cross-dimension architecture, shown in Fig.~\ref{fig:arch_b}; where from each dimension-projection we predict the clusters obtained from the other.

As the number of clusters, $k$, is decreased, the size of the neighbourhoods (points within a cluster) will increase and the approach will sacrifice local structure for global structure (since there is no constraint on how to position points within a cluster but only that they should be mapped to the same cluster). For example, if we assume that we produce a very high number of clusters such that on average two points form a cluster in the high-D representation, then the 2D latent-space will try to preserve the nearest-neighbour. On the other hand, if we assume there are only several clusters, then the 2D latent-space will preserve a more global structure such as which rooms the samples originate from (but not their position in the room). To enforce structure at multiple scales, instead of working with just one set of clusters (per dimension) we extract and predict multiple cluster assignments of different sizes concurrently.

\textbf{Triplet-loss:} Training a model with only the cluster-loss would generate a 2D representation that has good separation (i.e. points that are far apart in true-space are unlikely to overlap). However, it is unlikely to be topologically representative of the true environment, since a cross-entropy loss trained for softmax classification will try to encode different classes as orthogonal rays when projected to 2D.

Hence, we introduce a triplet-margin loss \eqref{equation:triplet} that operates directly on our 2D-space and pulls the representation together. Within a batch, let $x_i$ be an anchor point, sampled at timestamp $t_i$, we associate this with a positive $x_j$ and a negative $x_k$ so that the timestamp $t_i$ is closer to $t_j$ than to $t_k$, which come from the set of timestamps $T$. Let the function $d(x, x') = \| f_{\theta}(x) - f_{\theta}(x')\|^2$ be a measure of the Euclidean distance between the representations, and $M_t$ represent the margin: the minimum gap between the two distances that would reduce the loss to zero. Then, we define the loss $\mathcal{L_T}$ as:

\begin{equation} 
\label{equation:triplet}
\mathcal{L_T}=\frac 1N \sum_{(i,j,k) \in \mathcal T} \max(0, d(x_i, x_j) - d(x_i, x_k) + M_t)
\end{equation}

This loss incorporates the prior that within a small time-window (e.g a couple of seconds) there exists a linear relationship between time and distance.  For example; if we take one CSI sample, we would expect the distance to another CSI sample that is closer in time to also be closer in 2D-space. Assuming a small-enough window for this sampling, the distance assumption would hold even if the person does not walk in a straight-line (e.g. a circle).

The triplet-loss has two distinct advantages: (i) it helps pull together the separation created by the cross-entropy loss into a path, thus we observe smooth changes in latent space position even when the signal-strength abruptly changes (i.e. the person suddenly walks behind a pillar). (ii) The sampling strategy to determine positive and negative anchors uses the timestamp of the CSI sample, which introduces an extra source of information (the ordering of the packets) which is highly correlated with local spatial structure.

\subsection{Learning the Cartesian map (weakly-supervised)}

Assuming we have access to a handful of zone-level labels per zone, the 2D latent space produced in a self-supervised manner using the above method can be used to accurately perform zone-level positioning. This is shown in Table~\ref{tab:latent}. By incorporating additional real-world information into our model, the 2D-latent space can be transported to the real-world space by introducing two weakly-supervised losses. 

\textbf{Zone-loss:} We assume that a rough floor-plan is provided by the user where each zone is represented by a (bounding) box on a Cartesian map. Using a small-number of zone-level labels and the latent-space representation, we perform a KNN lookup to generate a predicted zone for every CSI sample. We then perform a key-value lookup on the floor-plan which is defined as $[B_{zone}]=([x_0,y_0],[x_1,y_1])$ i.e. a bounding-box per zone, to retrieve the bounding-box $B$ for the predicted zone. The loss $\mathcal{L_Z}$ then equates to the Manhattan distance $d_m(x,x')$ between the point and the box, if the point is predicted to be outside, and zero otherwise. 

\begin{equation} 
\label{equation:zone_loss}
\mathcal{L_Z}=\frac 1N \sum_{i=1}^N \max{(0,d_m(x_i, B_i))}
\end{equation}

\textbf{Access-Point Loss} It is well known that the power of a wireless signal decays exponentially as the distance from the source increases. 
We introduce this into our model by assuming that the precise location of the transmitter and receivers is provided by the user and create an access-point loss. $\mathcal{L_A}$, that operates similar to the triplet-margin loss.

\begin{equation} 
\label{equation:bs}
\mathcal{L_A}=\frac 1N \sum_{(i,k) \in \mathcal T}\sum_{a \in \mathcal A} \max(0, d(x_i, a) - d(x_k, a) + M_a)
\end{equation}

For each CSI sample in a batch, we sample a negative-anchor packet that is far-enough in time to have a difference in power, but close enough that it lies in the same zone (so that it is a harder example). Then, with respect to each access-point $a$ from the set $A$ which contain the receiver(s) and transmitter(s), we calculate the difference in normalized power. If this difference is greater than a threshold value, then the packet with a higher power ($x_i$) should be closer to the respective host by a hyper-parameter $M_a$ than the packet with the lower power ($x_k$). 

Since this relationship will not always be true, we only use this loss to initialize the model for several epochs before turning it off. However, initializing with this loss helps the model predict the correct orientation for some rooms. We observe that if a room is symmetric (and isolated) then the model has difficulty producing the correct orientation unless we include the access-point loss.

\subsection{Combined Loss}

The neural-network is trained end-to-end and the final loss term $\mathcal{L}$ is a combination of the two unsupervised losses and two weakly-supervised losses:

\begin{equation}
\mathcal{L} =\mathcal{L_C} + \mathcal{L_T} + \mathcal{L_Z} + \mathcal{L_A}
\label{equation:main}
\end{equation}

The losses can be summed with equal weights, and we omit an ablation that shows they are robust to small deviations in weights for brevity. However, the loss $L_{A}$ is only used to initialize the model for several epochs before it is turned off.

\section{Implementation Details}

\paragraph{Data collection} We generate the data-set by using multiple commercial IEEE 802.11 access points operating in the 5GHz band, and deploy them in test environments as shown in Fig.~\ref{fig:room} and Fig.~\ref{fig:room_ams}. Please note that zones are separated by drywall, glass, tall storage cabinets or concrete walls, the access points are in different zones, and several detection areas have no access points in line-of-sight. 
Each of the three receivers uses 8 dipole antennas arranged as a uniform circular array of 4cm radius. The transmitter uses only one of the antennas for transmission. 80MHz BW is used for the transmissions.
The CSI is estimated based on the reception of standard WiFi ACK packets, which the transmitter sends at periodic 10ms intervals. The CSI represents the channel between the transmitter antenna and each of its 8 receiver antennas, across 208 frequency tones that span the transmission bandwidth. Hence, the CSI is represented as a tensor of complex numbers per each packet $\in \mathbb{C}^{8\times 1\times208}$. We process the CSI and use only the magnitude, so that the input to the model uses real numbers. 

\paragraph{Architecture} We use a 2D architecture, ResNet-18 \cite{resnet}, as the convnet shown in Fig.~\ref{fig:arch_a} and Fig.~\ref{fig:arch_b}.
The network outputs a 128D feature vector and this is projected down to 2D with just a ReLU activation and one linear-layer. To predict the cluster assignments from the 2D features, we attach a MLP with one hidden-layer [2, 16, 64].

We extract a plurality of clusters from both feature-vectors: $k=\{256, 128, 64\}$ from the 128D features and $k=\{16, 8, 4\}$ from the 2D features. Incorporating more $k$s is helpful, however increases compute time. We set the margins $M_T$ and $M_A$ to 1 meter and verify this works well empirically.

\paragraph{Optimization} We use the SGD optimizer with weight-decay and a cosine decay learning-rate schedule (no restarts) and train for 100 epochs. The initial learning rate is 1e-2 which decays towards 1e-3. Our batch-size is 33, which means we sample: 11 anchors, 11 positive, and 11 negative points. The positive samples are within 2 seconds of the anchor and the negative points are within 2 to 4 seconds of the anchor. 

\paragraph{KNN} We use the faiss\cite{faiss} library for both clustering and KNN lookup. When reporting the latent-space quality metrics in Table~\ref{tab:latent}, we set $k=3$ and we use 42 locations as labels (for over 25,000 CSI samples).

\section{Experiments}

We present results on custom datasets which we have collected and annotated across three different environments (two offices and one house). We train our model on data collected on one day and evaluate on unseen-data collected on a different day.

\subsection{2D Office Environment No. 1}

\begin{figure}[htbp]
\centerline{\includegraphics[width=0.5\linewidth]{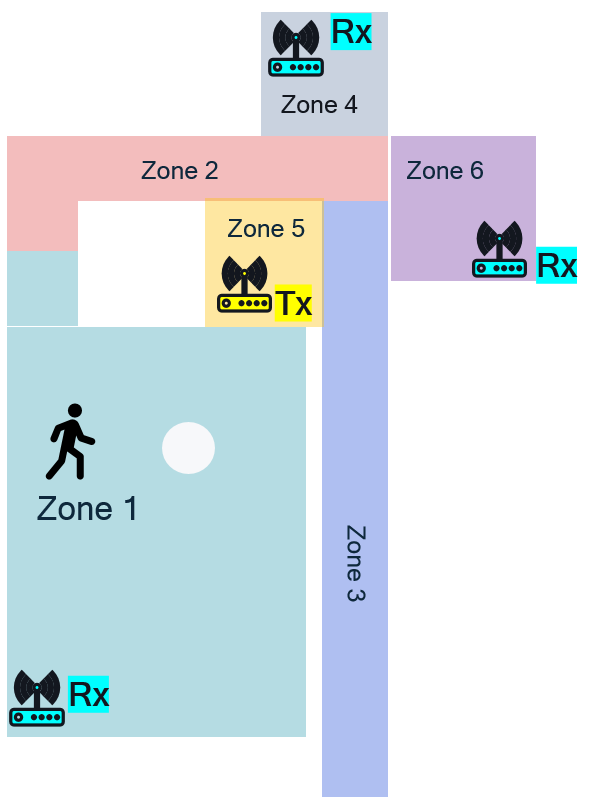}}
\caption{\textbf{Floor plan}: 14m x 20m. The white circle in Zone-1 represents a large pillar, which was part of the room structure.}
\label{fig:room}
\end{figure}

We train all models on data collected from one-day, this covers six zones in an indoor environment illustrated in Fig.~\ref{fig:room}, which total 885 seconds (almost 15 minutes). We report results for the same (train) dataset and also an unseen (test) dataset collected eight weeks earlier (with a similar length), to mimic real-world deployment where the user would expect the model to be functional without constant re-training.

\begin{table}[htbp]
\caption{Latent-space Quality}
\begin{center}
\begin{tabular}{|l|c|c|c|}
\hline
\textbf{Model}&\multicolumn{2}{|c|}{\textbf{Metrics}} \\
\cline{2-3} 
\textbf{} & \textbf{\textit{Kruskal stress}}& \textbf{\textit{KNN accuracy}} \\
\hline
PCA &0.56&0.84\\
\hline
t-SNE &0.77&0.55\\
\hline
UMAP &0.47 &0.86\\
\hline
Motion Metric$^{\mathrm{a}}$ &-&0.77\\
\hline
Triplet Loss (only) &0.47 &0.47 \\
\hline
Cluster Loss (only) &0.65 &0.73\\
\hline
WiCluster (unsupervised) &0.38 &\textbf{0.99}\\
\hline
WiCluster (weakly-supervised) &\textbf{0.15} &0.97\\
\hline
\end{tabular}
\label{tab:latent}
\end{center}
{$^{\mathrm{a}}$This approach uses the L2-norm of the CSI-sample to get an estimate of the signal-strength, and thus reduces the CSI tensor to a scalar that can be used for a KNN-lookup.}
\end{table}

\textbf{Zone-level Classification}. Table~\ref{tab:latent} shows the results for zone-level classification when we train a KNN classifier using 42 zone-level labels. As an additional assessment of latent-space quality we report the Kruskal stress \cite{kruskal} (following \cite{WirelessChannelCharting}), which measures the fidelity of all pairwise distances. We can observe that the cluster-loss achieves good separation and helps with global structure since it has a relatively high KNN accuracy. However, the triplet-loss has a better Kruskal stress metric but poor global separation. The combination of the two approaches: WiCluster (unsupervised), demonstrates the losses complement each other and result in a much lower stress and higher classification accuracy than any of the individual losses, or competing (unsupervised) methods. 

\begin{table}[htbp]
\caption{Precise Positioning in Office 1 \\ (Mean Error, in Meters)}
\begin{center}
\begin{tabular}{|c|c|c|c|c|}
\hline
\textbf{Zone no.}&\multicolumn{2}{|c|}{\textbf{WiCluster}}&\multicolumn{2}{|c|}{\textbf{Supervised}} \\
\cline{2-5} 
\textbf{} & \textbf{\textit{Train}}& \textbf{\textit{Test}}& \textbf{\textit{Train}}& \textbf{\textit{Test}} \\
\hline
1& 1.21& 1.41& 0.22& 1.83\\
\hline
2& 0.49& 0.80& 0.36& 2.32\\
\hline
3& 1.34& 1.91& 0.39& 1.17\\
\hline
4& 0.56& 0.73& 0.10& 0.69\\
\hline
5& 1.10& 1.24& 0.10& 0.57\\
\hline
6& 0.70& 1.63& 0.11& 0.99\\
\hline
\textbf{\textit{average}} & 0.90& 1.29& 0.22& 1.26\\
\hline
\end{tabular}
\label{tab:real_wicluster}
\end{center}
\end{table}

\textbf{Precise Positioning}. For Table~\ref{tab:real_wicluster} we use our weakly-supervised model to perform precise positioning. We report the mean error in meters for each zone (illustrated in Fig.~\ref{fig:room}) and the final average is the mean across the zones. The results show that although our approach has a higher error than the supervised model on the train-set (where the supervised model was shown all the precise-position labels), we are comparable on the test data-set. This suggests that we learn a representation that can generalize better, perhaps due to the self-supervised objectives.

\textbf{Robustness}. In Table~\ref{tab:robustness_2d} we report results on an additional 2D data-set, collected in the same environment, to investigate what happens when the user's behaviour changes. We consider the following test cases: regular walk (control), run, slow walk, carrying a large metal sheet, and a different test subject (with a different build). Although the largest disruption occurs when the subject carries a large metal-sheet with them, the average mean-zone error only increases by 13cm relative to the regular-walk.

\begin{table}[htbp]
\caption{Robustness in Office 1 \\ (Mean Error in Meters)}
\begin{center}
\begin{tabular}{|c|c|c|c|c|c|}
\hline
\textbf{Zone}&\multicolumn{5}{|c|}{\textbf{WiCluster}}\\
\cline{2-6} 
\textbf{} & \textbf{\textit{regular}}& \textbf{\textit{run}}& \textbf{\textit{slow walk}}& \textbf{\textit{metal}}& \textbf{\textit{subject}}\\
\hline
1& 1.49& 1.41& 1.61& 1.96& 1.30\\
\hline
2& 1.12& 1.25& 1.71& 1.46& 1.31\\
\hline
3& 1.55& 1.73& 1.52& 1.81& 1.38\\
\hline
4& 0.85& 0.74& 0.88& 0.65& 0.84\\
\hline
5& 1.17& 1.18& 1.31& 1.49& 1.39\\
\hline
6& 0.88& 0.57& 0.81& 0.45& 0.97\\
\hline
\textbf{\textit{average}} & 1.18& 1.15& 1.30& 1.31& 1.20\\
\hline
\end{tabular}
\label{tab:robustness_2d}
\end{center}
\end{table}

\subsection{2D Office Environment No. 2}

\begin{figure}[htbp]
\centerline{\includegraphics[width=0.5\linewidth]{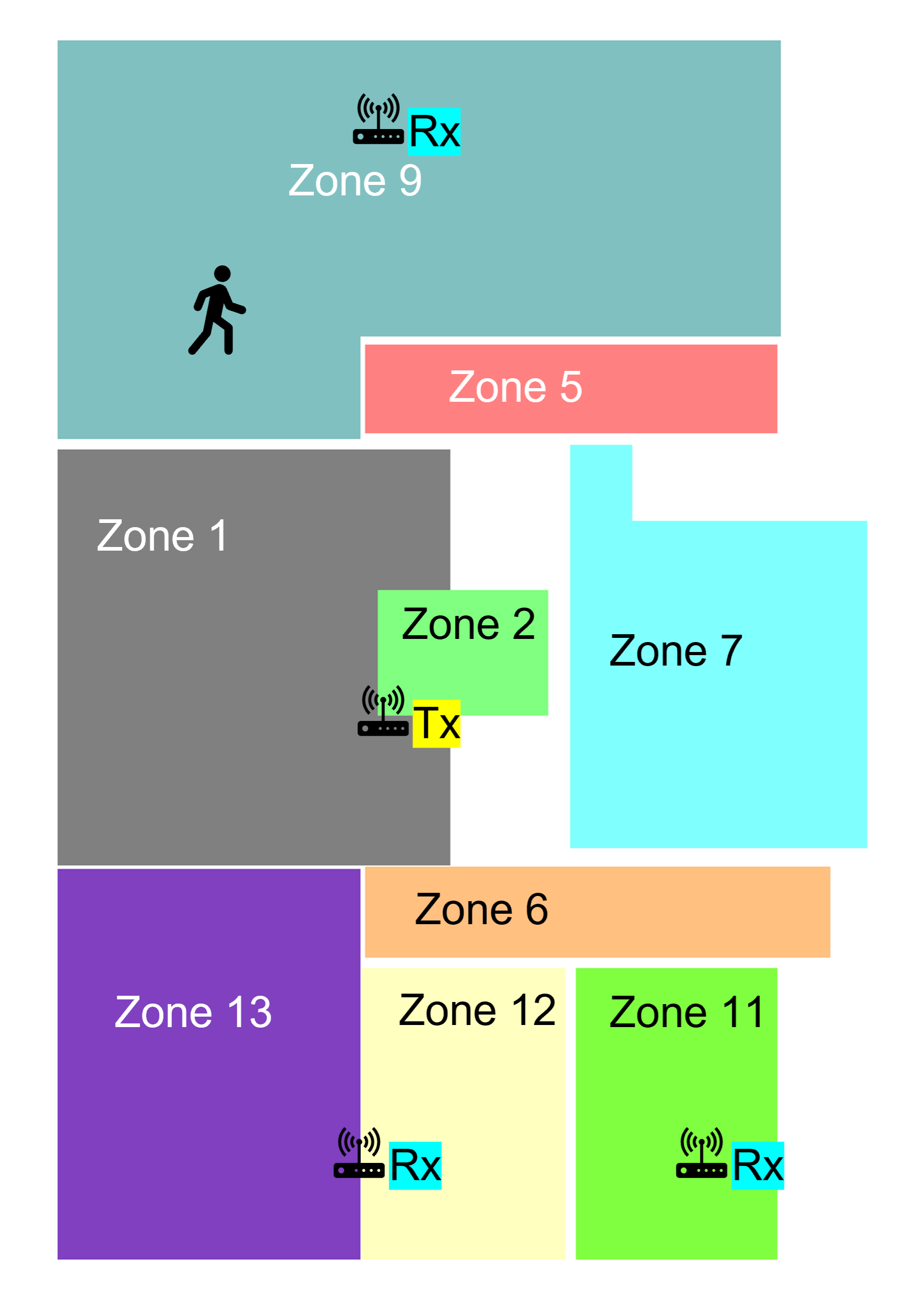}}
\caption{\textbf{Floor plan}: 15m x 21m. Zone 7 is lined with concrete}
\label{fig:room_ams}
\end{figure}

\begin{figure}[htbp]
\centerline{\includegraphics[width=1.0\linewidth]{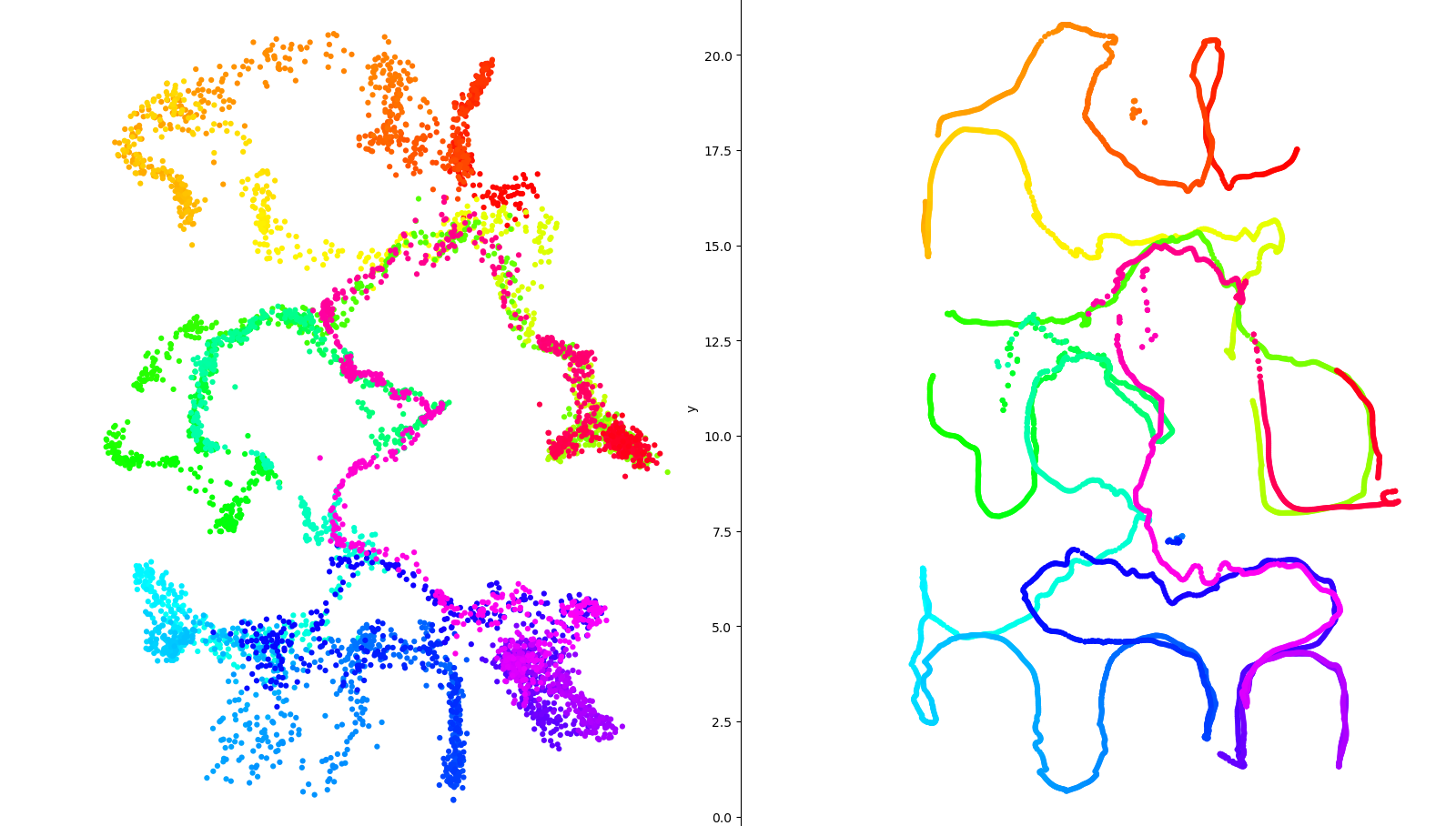}}
\caption{\textbf{Predicted 2D-positions} for a second office-environment. Prediction (left) vs Ground-Truth Locations (right). Color-scale corresponds to the timestamp of the CSI samples. Scale is in meters. }
\label{fig:ams}
\end{figure}

We demonstrate our approach on a second office environment that is larger and contains 9 zones. The floorplan is shown in Fig.~\ref{fig:room_ams} and we report our results in Table~\ref{tab:ams_wicluster} along with the corresponding visual of the predicted locations shown in Fig.~\ref{fig:ams}. Most of the precise positioning errors are under one meter, apart from those for Zone 7 which represents a room lined with concrete that severely dampens the signal. Similar to the robustness results in Table~\ref{tab:robustness_2d} we verify that there is very little difference if the person runs rather than walks (despite the algorithm being trained only on walking).

\begin{table}[htbp]
\caption{Precise Positioning in Office 2 \\ (Mean error, in Meters)}
\begin{center}
\begin{tabular}{|c|c|c|c|}
\hline
\textbf{Zone no.}&\multicolumn{3}{|c|}{\textbf{WiCluster}} \\
\cline{2-4} 
\textbf{} & \textbf{\textit{Train - walk}}& \textbf{\textit{Test - walk}}& \textbf{\textit{Test - run}} \\
\hline
1& 0.91& 0.81& 0.74\\
\hline
2& 0.82& 0.85& 0.91\\
\hline
5& 0.89& 1.40& 1.29\\
\hline
6& 1.04& 1.12& 1.31\\
\hline
7& 1.99& 2.26& 2.08\\
\hline
9& 1.08& 1.29& 1.24\\
\hline
11& 1.05& 1.24& 1.28\\
\hline
12& 1.29& 1.30& 1.26\\
\hline
13& 1.13& 1.05& 1.27\\
\hline
\textbf{\textit{average}} & 1.13& 1.26& 1.26\\
\hline
\end{tabular}
\label{tab:ams_wicluster}
\end{center}
\end{table}

\subsection{3D Home Environment}

\begin{figure}[htbp]
\centerline{\includegraphics[width=1.0\linewidth]{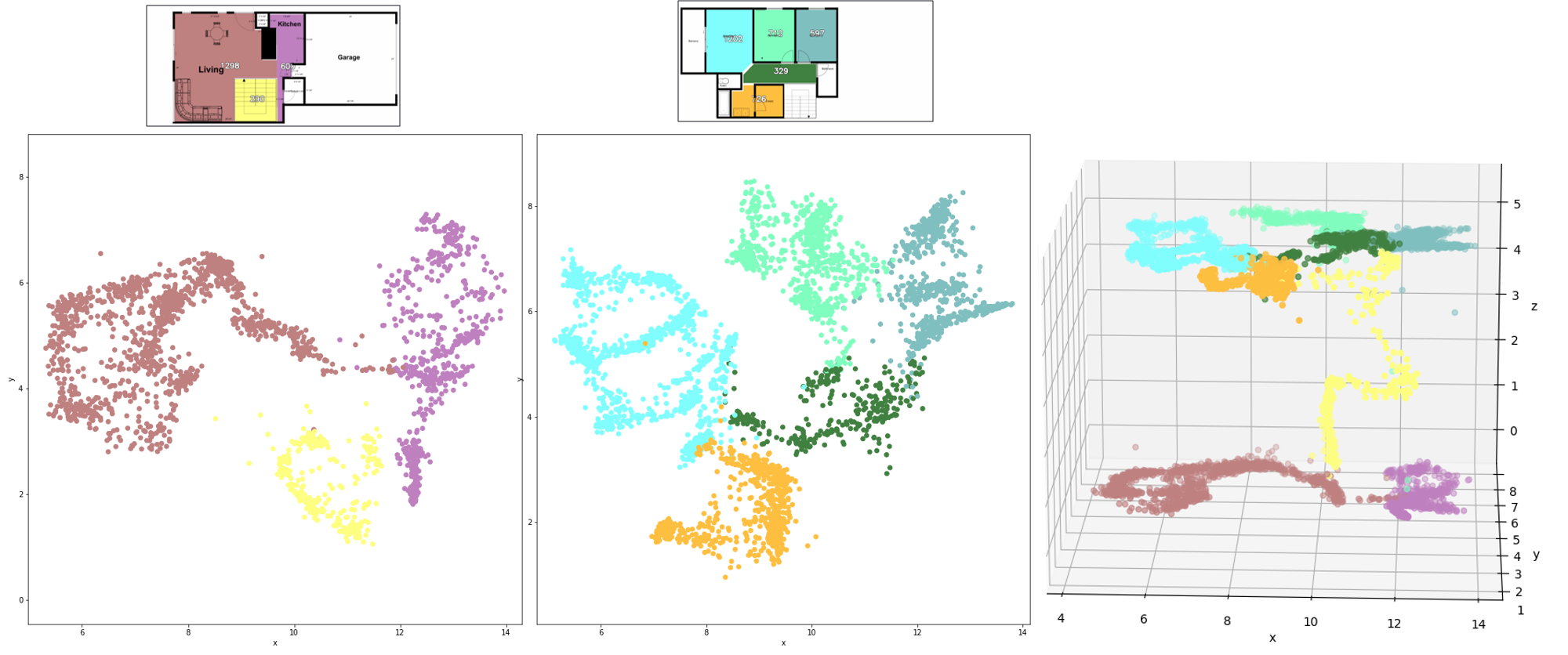}}
\caption{\textbf{Predicted 3D-positions} for a two-floor house. 3D-view (right) shows that stairs (yellow) have a continuous 3D position which connects the first (left) and second (middle) floors.}
\label{fig:home}
\end{figure}

To apply WiCluster to a two-story house, we deploy three access-points: one on the first floor, two on the second floor. The position along the z-axis must also be predicted, hence we change the latent space dimensionality from two to three. The z-axis is introduced into the Zone-loss through an L2 penalty. Although our approach predicts precise 3D positions, we do not have the corresponding ground-truth for this privacy-sensitive environment and can only evaluate the zone-level labels, shown in Table~\ref{tab:home}, however we still show the visualisation in Fig.~\ref{fig:home} since we can at-least verify the z-axis predictions.

\begin{table}[htbp]
\caption{3D Home Zone Classification Accuracy}
\begin{center}
\begin{tabular}{|c|c|c|c|c|}
\hline
\textbf{Zone}&\multicolumn{2}{|c|}{\textbf{WiCluster}}&\multicolumn{2}{|c|}{\textbf{Supervised}} \\
\cline{2-5} 
\textbf{} & \textbf{\textit{Train}}& \textbf{\textit{Test}}& \textbf{\textit{Train}}& \textbf{\textit{Test}} \\
\hline
Kitchen &1.00 &0.99 &1.00 &0.91\\
\hline
Living-room &1.00 &0.94 &1.00 &0.99\\
\hline
Stairs &0.95 &0.94 &0.99 &0.97\\
\hline
Hallway &0.94 &0.95 &0.99 &0.98\\
\hline
Bedroom-1 &0.97 &0.92 &0.99 &0.99\\
\hline
Bedroom-2  &0.96 &0.98 &1.00 &0.97\\
\hline
Bedroom-3  &0.99 &0.97 &1.00 &0.95\\
\hline
Closet  &1.00 &1.00 &1.00 &0.99\\
\hline
\textit{average} &0.98 &0.96 &0.99 &0.97\\
\hline
\end{tabular}
\label{tab:home}
\end{center}
\end{table}

\section{Conclusion}

We propose a novel dimensionality reduction technique that uses cross-dimension and multi-scale clustering to preserve local and global structure. Combining this with representation learning we can generate a 2D/3D latent-space that is topologically close to the real space and can be used for zone-level classification already. We demonstrate that by incorporating some real-world measurements we can transport this to a Cartesian map. As a result we are able to predict precise 2D or 3D positions for a single person from Wi-Fi CSI data, without using any precise position labels during training. We demonstrate that our system achieves meter-level accuracy in three different realistic environments: two offices and one multi-story home. The positioning system performs well even in rooms without a line-of-sight connection to either the transmitter or receiver.

\smallskip
\textbf{Acknowledgements} We thank Fatih Porikli, Hanno Ackermann and Berkay Kicanaoglu for their feedback and discussions.

\vspace{12pt}

\end{document}